\documentstyle[12pt]{article}

\begin{document}
\title{Semiclassical Effects Induced by Aharonov-Bohm 
Interaction Between a Cosmic String and a Scalar Field}
\author{M. E. X. Guimar\~aes \\
\mbox{\small{Department of Physics and Astronomy }}\\
\mbox{\small{University of Wales, College of Cardiff}} \\
\mbox{\small{PO Box 913, Cardiff CF2 3YB, UK}}}
\date{14 February 1997}
\maketitle
\begin{abstract} 
In the context of the vacuum polarization effect, 
we consider the 
backreaction of the energy-momentum tensor of a charged 
scalar field on the background metric of a cosmic string carrying a 
magnetic flux $\Phi$. 
Working within the semiclassical 
approach to the Einstein eqs. we find the first-order (in $\hbar$) metric 
associated to the magnetic flux cosmic string. 
We show that the contribution to the vacuum 
polarization effect coming from the Aharonov-Bohm interaction is larger than 
the one coming from the non-trivial gravitational interaction. 
\newline
{\em Classification PACS:} 81T20, 83C47
\end{abstract} 

\section*{Introduction}

In General Relativity, a 
static, straight axially symmetric cosmic string is described by the 
metric \cite{go}
\begin{equation}
ds^2 = -dt^2 +dz^2 +d\rho^2 +B^2\rho^2d\varphi^2
\end{equation}
in cylindrical coordinates $(t,z,\rho,\varphi)$ such that $\rho \geq 0$ and 
$0 \leq \varphi < 2\pi$. The constant $B$ is related to the linear mass 
density $\mu$ of the string\footnote{Throughout this paper, we 
work in the system of units in which 
$G=c=1$ and $\hbar\sim 2.612 \times 10^{-66}$ cm$^2$.}: $B= 1-4\mu$. 
For GUT strings, $\mu$ is of order $\mu \sim 10^{22}$ g/cm. 

Metric (1) is locally but not globally flat. The presence of the 
string 
leads to an azymuthal deficit angle $\Delta = 8\pi\mu$ and, as a result, this 
spacetime has a conical singularity \cite{vi}. 
One of the most interesting features of 
this spacetime is that fields and particles are sensitive to its global 
structure and physical effects may arise due solely to the global properties 
of this metric. One of these effects - the vaccum polarization - has been 
extensively studied in the literature \cite{he,dow} and can 
be understood as an analog to the Casimir effect \cite{bi} in which the  
conducting planes 
here form an angle equal to the deficit angle $\Delta$. Then, a scalar field 
placed in the spacetime outside the string has its vacuum state 
polarized due to non-trivial periodicity conditions on the azymuthal 
angle $\varphi$ \cite{he}. In papers of reference \cite{dow} a more general 
situation 
has been considered in which the cosmic string carrying a magnetic flux 
$\Phi$ interacts  
with a charged scalar field placed in the metric (1). In this case, the 
vacuum polarization arises not only via non-trivial gravitational 
interaction (i.e, the global conical structure) but also via 
Aharonov-Bohm (AB) interaction:  
The scalar field acquires an additional phase shift proportional to 
the magnetic flux in spite of the fact that it is 
placed in a region where there is no magnetic 
field \cite{se}. This situation is a realization in Cosmology of 
the original AB effect. In papers \cite{dow} the 
non-vanishing vacuum expectation values (VEV) of the energy-momentum tensor 
for the scalar field in the fixed background (1) were determined. 
However, if we want to compute the contribution of both the magnetic flux 
and the scalar field on the original metric of the cosmic string, 
this non-vanishing energy-momentum tensor must be 
taken into account to determine a more realistic 
spacetime metric associated with the 
magnetic flux cosmic string. This is the purpose of the present letter. 

Throughout this paper we will work in the so-called semiclassical approach 
to 
the Einstein eqs. $G_{\mu\nu}=8\pi \langle T_{\mu\nu} \rangle $ and we will 
treat this problem using the perturbative approach as in Hiscock's paper 
\cite{hi}. In this approach, the first-order (in $\hbar$) $\langle 
T_{\mu\nu} \rangle$ is treated as a matter perturbation of the zeroth-order 
metric (1) and can be used to compute the first-order metric 
perturbation associated to it by solving the linearized Einstein's eqs. 
about 
the zeroth-order metric. In the present case, there will be contributions 
from 
both the non-trivial gravitational and the AB interactions. 
We find the gravitational force associated with the backreaction of the 
$\langle T_{\mu\nu}\rangle$ and the first-order corrections to the deficit 
angle. Our main 
result is that the AB is the leading interaction between the 
magnetic flux cosmic string and the charged scalar field and 
dominates over the gravitational 
interaction. 
That is, the sign of both the gravitational force and the deficit angle is 
determined by the AB interaction.

\section*{Semiclassical Effects Induced by the Backreaction of the $\langle 
T_{\mu\nu} \rangle$} 

The $\langle T^{\mu}_{\nu}\rangle$ for a massless, charged scalar field in 
the geometry (1) is \cite{dow} 
\begin{eqnarray}
\langle T^{\mu}_{\nu} \rangle  & = & \frac{\hbar}{\rho^4}\left[ 
\omega_4(\gamma) - 
\frac{1}{3}\omega_2(\gamma) \right] diag (1,1,1,-3) \nonumber \\
& & +4 \left(\xi -\frac{1}{6}\right)\frac{\hbar}{\rho^4}\omega_2(\gamma) 
diag (1,1,-1/2,3/2) ,
\end{eqnarray}
where the constants $\omega_2(\gamma)$ and $\omega_4(\gamma)$ are given 
by the following expressions
\[
\omega_2(\gamma)  =  -\frac{1}{8\pi^2} \left[
\frac{1}{3}-\frac{1}{2B^2}
[ 4(\gamma-\frac{1}{2})^2 -\frac{1}{3} ] \right] 
\]
\begin{eqnarray}
\omega_4(\gamma) & = & - \frac{1}{720\pi^2} [ 11 -\frac{15}{B^2}
[ 4(\gamma-\frac{1}{2})^2-\frac{1}{3}]  \nonumber \\
& & +\frac{15}{8B^4}[ 16(\gamma-\frac{1}{2})^4
-8(\gamma - \frac{1}{2})^2+\frac{7}{15} ]  ] , 
\end{eqnarray}
valid only\footnote{This {\em mathematical} restriction 
arises from successive integrations to obtain the $\langle T^{\mu}_{\nu}
\rangle$. 
However, it does not correspond to a {\em physical} restriction because 
strings of cosmological interest are of order $\mu \sim 10^{-6}$ (recall 
$B=1-4\mu$). For more details, see, for instance, Guimar\~aes and Linet 
\cite{dow}.} when $B>1/2$  
and $\gamma$ is the fractional part of $\{ \frac{\Phi}
{\Phi_0} \}$, $\Phi_0$ being the quantum flux $\Phi_0 =2\pi\hbar/e$ and lies 
in the domain $0 \leq \gamma < 1$. 
Let us now define the dimensionless quantities
\begin{eqnarray}
A(\gamma) & \equiv & \omega_4(\gamma) -\frac{1}{3}\omega_2(\gamma) 
\nonumber \\
B(\gamma) & \equiv & 4(\xi - \frac{1}{6})\omega_2(\gamma) ,
\end{eqnarray}
in such a way that the components of $ \langle T^{\mu}_{\nu} \rangle $ 
can now be rewritten as 
\begin{eqnarray}
\langle T^t_t \rangle  & = & \langle T^z_z \rangle = \frac{\hbar}{\rho^4}[
A(\gamma)+B(\gamma)] , \nonumber \\
\langle T^{\rho}_{\rho}\rangle  & = & \frac{\hbar}{\rho^4}[A(\gamma)-
\frac{1}{2}B(\gamma)] ,  
\nonumber \\
\langle T^{\varphi}_{\varphi}\rangle  & = & -3 
\frac{\hbar}{\rho^4}[A(\gamma)-\frac{1}{2}
B(\gamma)] .
\end{eqnarray}
The $\langle T^{\mu}_{\nu}\rangle $ above is linear in $\hbar$ and its 
dimensionality is 
$[L]^{-2}$. We can now attempt to solve the semiclassical Einstein's 
equations $G_{\mu\nu}=8\pi\langle T_{\mu\nu}\rangle $ 
at linearized level to obtain the 
first-order metric perturbation associated to the backreaction of the 
$\langle T^{\mu}_{\nu}\rangle $ (5). We follow here the same approach as 
Hiscock in 
paper \cite{hi} and we will later compare our results (in which the magnetic 
flux is present) with his results (in which there is no magnetic flux).

Following Hiscock's procedure \cite{hi}, we set a static, 
cylindrically symmetric metric in the general form
\begin{equation}
ds^2=e^{2\Phi(\rho)}(-dt^2+dz^2+d\rho^2)+e^{2\Psi(\rho)}d\varphi^2,
\end{equation}
where $\Phi$ and $\Psi$ are functions of $\rho$ only; and we expand this 
metric about the background metric
\begin{equation}
\Phi=\phi_0+\phi \;\;\; and \;\;\; \Psi= \psi_0+\psi
\end{equation}
where, for metric (1), we have $\phi_0=0$ and $\psi_0=\ln(B\rho)$. Therefore, 
we obtain the linearized Einstein's equations with source (5)
\[
\psi''+\phi''+\frac{2}{\rho}\psi'=8\pi[A(\gamma)+B(\gamma)]\hbar\rho^{-4}
\]
\[
\frac{2}{\rho}\phi'=8\pi[A(\gamma)-\frac{1}{2}B(\gamma)]\hbar\rho^{-4} 
\]
\begin{equation}
2\phi''=-24\pi[A(\gamma)-\frac{1}{2}B(\gamma)]\hbar\rho^{-4} .
\end{equation}
The general solutions for eqs. (8) can be easily found
\begin{eqnarray}
\phi & = & -2\pi[A(\gamma)-\frac{1}{2}B(\gamma)]\hbar\rho^{-2}+C ,
\nonumber \\
\psi & = & 10\pi[A(\gamma)+\frac{1}{10}B(\gamma)]\hbar\rho^{-2} 
+D\rho^{-1} +E .
\end{eqnarray}
The exterior metric (corrected at first-order in $\hbar$) of the magnetic 
flux cosmic string is then obtained 
\begin{eqnarray*}
ds^2 & = & \left[ 1 -4\pi\frac{\hbar}{\rho^2}[A(\gamma)-\frac{1}{2}B(\gamma)]
\right] (-dt^2+dz^2+d\rho^2) \nonumber \\
& & +\left[ 1 +20\pi\frac{\hbar}{\rho^2}[A(\gamma)+\frac{1}{10}B(\gamma)]
\right](1-4\mu)^2\rho^2d\varphi^2 .
\end{eqnarray*} 

The semiclassical approach is legitimated so long as the first-order 
perturbations are small compared to one: 

\[
\mid \hbar \rho^{-2} [A+\frac{1}{10}B]\mid \ll 10^{-2} , 
\]
which means that $ \rho \gg 10 [\hbar (A+\frac{1}{10}B)]^{1/2}$. The 
rhs is approximately equal to $\approx (\hbar\mu)^{1/2}$. Since the 
radius of a physical cosmic string is approximately of the same 
order\footnote{Of the order of the Compton wavelength of the Higgs bosons 
involved in the phase transitions leading to the formation of the cosmic 
string, $\sim 10^{-30}$ cm.}, 
this means that $\rho \gg \rho_s $ . That is, the semiclassical approach 
is valid everywhere outside the cosmic string. 

Now, if we want to describe the string in a coordinate system such that the 
radial coordinate measures the proper radius from the string, we make the 
change of variables
\[
r=\rho + 2\pi\frac{\hbar}{\rho}[A(\gamma)-\frac{1}{2}B(\gamma)] ,
\]
such that $g_{rr}=1$ and therefore the exterior metric becomes
\begin{eqnarray}
ds^2 & = & \left[ 1-4\pi\frac{\hbar}{r^2}[A(\gamma)-\frac{1}{2}B(\gamma)] 
\right](-dt^2+dz^2)+dr^2 \nonumber \\
& & +(1-4\mu)^2r^2\left[ 1+16\pi\frac{\hbar}{r^2}
[A(\gamma)+\frac{1}{4}B(\gamma)] \right] d\varphi^2 .
\end{eqnarray}
The geometry of the $(r,\varphi)$-space is no longer flat, 
but asymptotically approaches the zeroth-order metric (1). 
The first consequence is the 
appearance of a non vanishing gravitational force on a massive test particle.
If we set $g_{00}= - [1+2\Phi]$ where $\Phi$ is the newtonian potential, 
we have
\[
f^r =-4\pi{\hbar}{r^3}[A(\gamma)-\frac{1}{2}B(\gamma)] .
\]
Using the definitions (4), we obtain a general expression for the 
gravitational force
\begin{equation}
f^r=-4\pi\frac{\hbar}{r^3}[\omega_4(\gamma)-\frac{1}{3}\omega_2(\gamma)-
2(\xi-\frac{1}{6})\omega_2(\gamma)] .
\end{equation}
For the particular values $\xi =0$ (minimal coupling) and 
$\xi =1/6$ (conformal parameter), the general expression (11) of the 
gravitational force takes the form  
\begin{equation}
f^r=-4\pi\frac{\hbar}{r^3}\omega_4(\gamma)
\end{equation}
and  
\begin{equation}
f^r=-4\pi\frac{\hbar}{r^3}[\omega_4(\gamma)-\frac{1}{3}\omega_2(\gamma)], 
\end{equation}
respectively. 

We can also obtain the first-order corrections to the deficit angle. Let us 
define it as $\Delta\varphi= 2\pi -C/r$, where $C$ is the circunference of 
a circle centered around the string at a fixed proper radius $r$ from it. 
Therefore, for metric (10), the deficit angle has the following 
expression (after using defintions (4))
\[
\Delta\varphi=8\pi\mu -(1-4\mu)16\pi^2\frac{\hbar}{r^2}[\omega_4(\gamma)-
\frac{1}{3}\omega_2(\gamma)+(\xi-1/6)\omega_2(\gamma)] .
\]
For the particular values $\xi=0$ and $\xi=1/6$ of the coupling parameter, 
this general expression reduces to
\begin{equation}
\Delta\varphi=8\pi\mu -(1-4\mu)16\pi^2\frac{\hbar}{r^2}[\omega_4(\gamma)
-\frac{1}{2}\omega_2(\gamma)], 
\end{equation}
and
\begin{equation}
\Delta\varphi=8\pi\mu -(1-4\mu)16\pi^2\frac{\hbar}{r^2}[\omega_4(\gamma) 
-\frac{1}{3}\omega_2(\gamma)],
\end{equation}
respectively.

\section*{Comparison with the Case $\gamma = 0$ and Concluding Remarks}

Now we are able to make some conclusions about the results obtained in the 
previous section. 
Let us first consider the gravitational force (11). 
In the case where there is no magnetic flux ($\gamma=0$) the gravitational 
force is always {\em attractice} for both minimal ($\xi =0$) and conformal 
($\xi =1/6$) couplings. However, this behaviour changes when the magnetic 
flux is present (and $\gamma$ lies in the domain $0 <\gamma <1$). Indeed, 
 it is easy to see from (12) and (13) that the gravitational force is 
{\em repulsive} for both minimal and conformal couplings. 

Considering now the deficit angle, again the behaviour changes if the 
magnetic flux is present or not. When it is absent, the deficit angle 
{\em increases} as $r\rightarrow 0$ for minimally coupled scalar field and 
{\em decreases} as $r\rightarrow 0$ for conformally coupled field. When 
the magnetic flux is present, and $\gamma$ lies in the domain $0<\gamma<1$, 
the result is different:  
The deficit angle {\em decreases} as $r\rightarrow 0$ for 
minimally  coupled scalar field and {\em increases} as 
$r\rightarrow 0$ for conformally coupled field.

It seems, thus, clear from these analysis that the sign of both the 
gravitational 
force and the deficit angle is determined by the AB interaction 
between the magnetic flux cosmic string and the charged scalar field. We can, 
then, conclude that the contribution coming from the AB interaction 
dominates over the one coming from the non-trivial gravitational interaction. 
It is interesting to remark that this result agrees with previous statement 
by Alford and Wilczek  (and further by de Souza Gerbert)
\cite{alf} although in a diferent context: 
In these papers the authors show that the cross section coming from 
AB scattering of fermions in presence of a cosmic string is much 
larger than the one coming from the gravitational scattering.

\section*{Acknowledgement}

It is a pleasure to thanks Prof. Bernard Linet for helpful discussions and 
a critical reading of a previous version of this manuscript.

\end{document}